# Broadband conversion of microwaves into propagating spin waves in patterned magnetic structures

F. B. Mushenok,[1] R. Dost,[2] C. S. Davies,[1] D. A. Allwood,[2] B. J. Inkson,[2] G. Hrkac,[1] and V. V. Kruglyak[1,*]

[1] *School of Physics, University of Exeter, Stocker road, Exeter, EX4 4QL, United Kingdom*

[2] *Department of Materials Science and Engineering, University of Sheffield, Sheffield, S1 3JD, United Kingdom*

## Abstract

We have used time-resolved scanning Kerr microscopy (TRSKM) and micromagnetic simulations to demonstrate that, when driven by spatially uniform microwave field, the edges of patterned magnetic samples represent both efficient and highly tunable sources of propagating spin waves. The excitation is due to the local enhancement of the resonance frequency induced by the non-uniform dynamic demagnetizing field generated by precessing magnetization aligned with the edges. Our findings represent a crucial step forward in the design of nanoscale spin-wave sources for magnonic architectures, and are also highly relevant to the understanding and interpretation of magnetization dynamics driven by spatially-uniform magnetic fields in patterned magnetic samples.

---

[*] Corresponding author: V.V.Kruglyak@exeter.ac.uk

The challenge of generation of short-wavelength spin waves hampers both the immediate and future development of magnonic technology.[1,2] Conventionally, spin waves are excited using miniaturized current-carrying elements, whereby the associated non-uniform microwave magnetic field injects propagating spin waves with finite wavelength into the nearby magnetic film.[3] By tuning the microwave frequency and dimensions of the microwave circuitry, one can control the frequency / wavelength of the emitted plane spin waves,[4] or even produce collimated spin-wave beams.[5] Alternatively, focused optical pulses,[6] spin-transfer torques[7] or parametric pumping[8] can be used to trigger spin waves. However, when implemented within miniaturized systems, these approaches suffer from substantial difficulties associated with the Joule heating (inherent to charge transport) and the spatial extent of the magnonic source (dictating the smallest wavelength that can be achieved).

An alternative scheme, suggested by Schlömann in 1964,[9] involves coupling between a uniform microwave magnetic field and propagating spin waves that is mediated by the non-uniform internal magnetic field. This non-uniformity gives rise to a graded distribution of the locally defined ferromagnetic resonance (FMR) frequency, $f_{FMR}(r)$, across the magnetic body. A driving magnetic field of frequency $f$ then couples resonantly (and therefore most efficiently) only to specific regions of magnetization with $f_{FMR} \sim f$. The non-uniform effective magnetic field means that translational symmetry is broken and that linear momentum therefore does not need to be conserved. This causes the locally resonating magnetization to launch spin waves of finite wavelength into the adjacent region of the sample. Thereby, this 'Schlömann' mechanism of spin-wave excitation removes the need for localization of the driving microwave field, which also renders obsolete construction of a complex array of electrical nano-circuitry.

The Schlömann mechanism has been extended to patterned microstructures in which the incident microwave field resonantly couples to one of the discrete spin-wave modes of an "antenna" formed by either semi-infinite patches or microstripes, either located above or patterned from the same magnetic film as the magnonic waveguide.[10-13] In Ref. 14, the use of a vortex core has been proposed to serve as such an antenna. Such samples can be classified as systems with Fano resonances,[15,16] created by the overlap of the discrete spectrum of the antenna and the continuous spectrum of the magnonic waveguide. In addition to the resonant excitation of spin waves, a peculiar reverse effect, in which the discrete resonance absorbs fully an incident propagating spin wave, could be observed and exploited in such systems.[17] However, the reliance on the resonance with discrete normal modes of well-defined parts of the sample means that the frequency tuning range of the emitted spin waves is limited by the resonance linewidth of the relevant normal mode and therefore by the magnetic damping of the antenna, with an obvious adverse effect on the device performance. This contrasts with the Schlömann mechanism, in which the resonating regions do not confine spin waves and therefore are not characterized by a discrete spectrum. Instead, the tuning is determined by the continuous distribution of $f_{FMR}(r)$, so that microwaves of different frequencies couple to slightly different sample regions.

Here, we use time-resolved scanning Kerr microscopy (TRSKM) to demonstrate that magnetization aligned along edges of patterned magnetic samples can serve as a source of propagating magnetostatic spin waves. Specifically, we consider a transversely magnetized Permalloy stripe excited by a spatially-uniform harmonic magnetic field at frequencies above that of the uniform FMR. The emission of spin waves propagating from the stripe edges is imaged experimentally and corroborated using micromagnetic simulations. To explain our findings, we have developed a hybrid numerical / analytical framework that enables $f_{FMR}(r)$ to be calculated across the stripe. Compared to the majority of previous studies in which the variation of $f_{FMR}(r)$ was due to the static field distribution,[9,18-20] our formalism takes into account both the static and dynamic contributions to the demagnetizing field. This calculation allows us to pinpoint regions of locally enhanced $f_{FMR}(r)$ and thereby the location of the resonantly excited magnetization and source of the observed highly tunable spin-wave excitation. The increase in $f_{FMR}(r)$ near the edges of patterned magnetic structures makes the demonstrated scheme of spin-wave generation truly general and relevant to the excitation of spin waves not only in stripes but also in any magnetic samples possessing non-uniformity, either geometrical and/or compositional. Our results will find application in interpretation of the magnetization dynamics observed in a variety of experimental schemes,[10,12-13,14,21-22,23,24,25] as well as being a crucial step towards development of miniaturized all-magnonic technology.

A Permalloy ($Ni_{80}Fe_{20}$) stripe of 100 nm thickness (along $z$ axis), 10 μm width and 300 μm length was fabricated on a 170 μm thick glass coverslip using electron beam lithography and thermal evaporation deposition. The sample was laid face-down onto the 500 μm wide signal line of a microwave coplanar waveguide, as shown in Fig. 1 (a). This geometry ensured an effectively uniform spatial distribution of the pumping magnetic field $h$ across the Permalloy stripe. Throughout all the experiments described here, the stripe was in-plane magnetized parallel to its short axis ($y$-direction) by a uniform bias magnetic field $H = 200$ Oe. The pumping magnetic field was created by the microwave current in the coplanar waveguide. The dynamic out-of-plane component of the excited magnetization was probed in the TRSKM via the polar Kerr effect. Further details of our TRSKM system are found in Refs. 10–13.

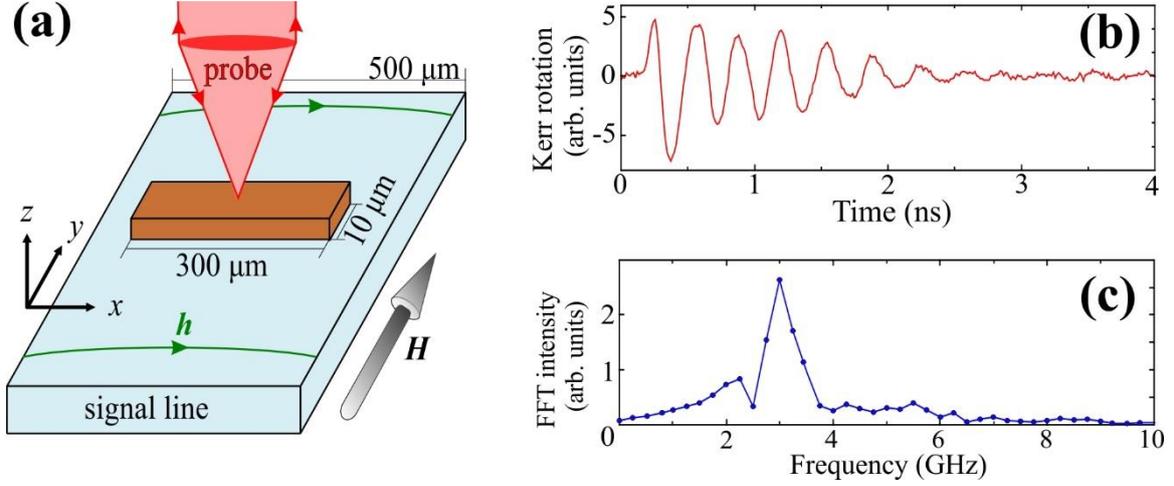

Fig. 1 (a) The studied Permalloy stripe is shown schematically with the coplanar waveguide used to deliver the microwave pumping field $h$. (b) A typical time-resolved Kerr signal acquired from the center of the stripe in response to its pulsed excitation is shown for the bias field $H = 200$ Oe. (c) The Fourier spectrum calculated from the signal in (b) is shown.

In the first stage of the experiments, the sample was excited by a short (~70 ps) pulsed magnetic field was used to excite broadband dynamics in the sample, and the time-resolved magnetic response was recorded by focusing the TRSKM's optical probe spot at a range of characteristic positions on the stripe. A typical time-resolved Kerr signal and its Fourier spectrum are shown in Figs. 1 (b) and (c), respectively. The dominant frequency peak observed at about 3 GHz corresponds to the quasi-uniform precessional mode spanning the majority of the stripe, while the smaller peak at 2.25 GHz is due to the edge mode.[21,24] In the second stage of experiments, $h$ had instead a harmonic temporal character and was tuned to a frequency *above* that of the quasi-uniform precession (i.e. > 3 GHz). Fig. 2 (a) – (h) present Kerr snapshots of the dynamic out-of-plane component of the magnetization in the stripe, taken in phase steps of $\pi/4$, i.e. in time steps of 22 ps at the excitation frequency of 6.24 GHz. A uniform background was subtracted from the images to emphasize the non-uniform dynamics.

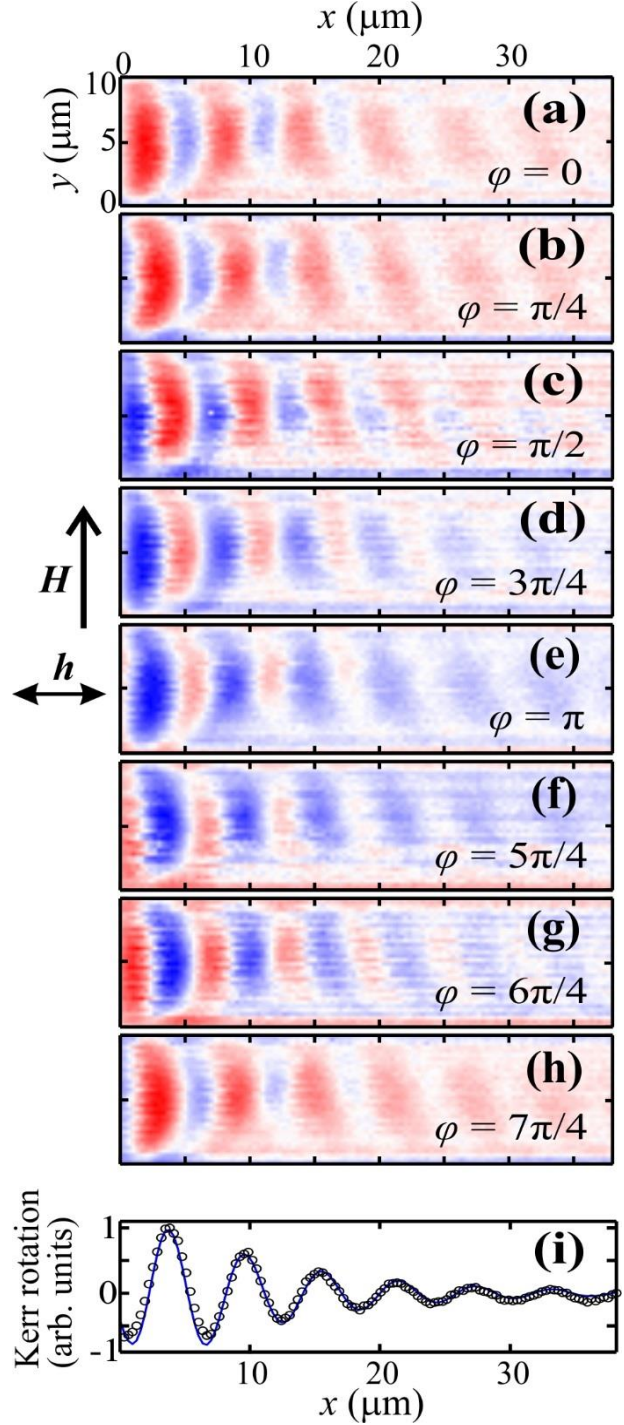

Fig. 2  (a)–(h) Kerr images of the dynamic magnetization are shown for the excitation at 6.24 GHz in phase $\varphi$ (time) steps of $\pi/4$ (22 ps). (i) The black circles and blue solid line show the Kerr signal from panel (b) averaged over the middle part of the stripe's width ($y = 5 \pm 2.5$ μm) and its fit to an exponentially decaying harmonic function, respectively.

The spatial distribution of dynamic magnetization in Fig. 2 reveals the propagating character of the excited spin waves. The decay of the spin-wave amplitude from left to right indicates that the spin-wave source is located close to the stripe's left edge. The tilt of the wave fronts is attributed to a modest canting of the static magnetization from the in-plane symmetry axis.[13] To deduce the wavelength λ and decay length *l* of the excited spin waves, the Kerr signals (over the middle part of the stripe's width) were fitted to an exponentially decaying harmonic function, as shown in Fig. 2 (i). At the frequency of 6.24 GHz, the fit yielded 6.1 μm and 10.8 μm for the spin-wave wavelength and decay length, respectively. The Kerr imaging was also repeated at frequencies of excitation ranging between 4 GHz and 8 GHz. The acquired snapshots of the dynamic out-of-plane components of magnetization (generally, corresponding to different relative phases between the driving field and precession) are shown in Fig. 3 (a) – (g). These results demonstrate that the scheme of excitation is efficient across a broad frequency range. Using the previously discussed fitting approach, the wavelength and decay length of the observed spin waves were extracted (see Supplementary information) and found to decrease monotonically as the excitation frequency increased. The deduced dispersion of the excited spin waves (Fig. 4 (d)) has a clear Damon-Eshbach character,[26] as expected from the implemented geometry.

Our experimental results agree well with those obtained from micromagnetic simulations, described in detail in Supplementary information. First, the spin-wave dispersion (Fig. 4 (d)) was calculated from simulations performed for an infinitely long stripe with cross-section of $10 \times 0.1$ μm$^2$ and with Permalloy-like magnetic parameters. The dominant dispersion branch was found to correspond to Damon-Eshbach spin waves, and the less intense branches of lower frequency were assigned to quantized spin waves confined within the stripe width. Then, micromagnetic simulations of a finite-length stripe were performed to confirm that propagating spin waves can indeed be driven by a uniform harmonic magnetic field. The wavelengths of the spin waves emitted from the ends of the stripe in this simulation set agreed with the dispersion curve calculated from simulations for an infinitely long stripe (Fig. 4 (d)).

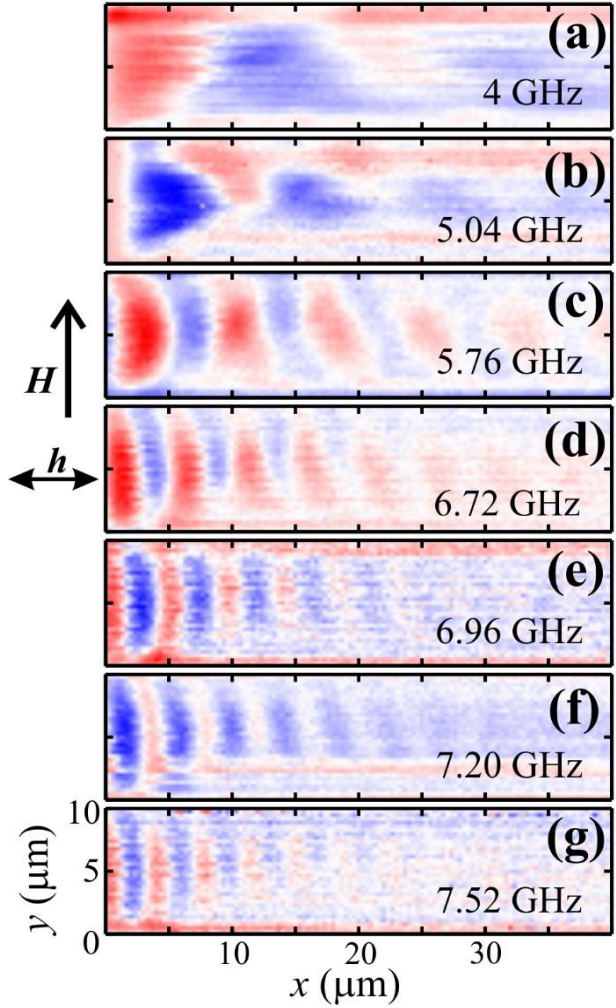

Fig. 3  (a)–(g) Kerr images are shown for the magnetic stripe driven at frequency of 4 GHz, 5.04 GHz, 5.76 GHz, 6.72 GHz, 6.96 GHz, 7.20 GHz, and 7.52 GHz, respectively.

The observed broadband spin-wave emission cannot be explained based on the static demagnetizing field distribution (as e.g. in Ref. 9) but the dynamic demagnetizing field needs to be invoked. For a uniformly magnetized sample, magnetic charges and the maxima and minima of the demagnetizing field (which can be static or dynamic in origin) occur at sample boundaries.[9,27] In particular, as one moves closer to the edge along which the sample is magnetized, the dynamic demagnetizing field increases in strength. This fact was used in Refs. 28–29 to obtain the boundary conditions characterizing the magnetization at the edges of stripes and therefore to calculate the spectrum of standing width modes. We argue that the same dynamic demagnetizing field is also responsible for the excitation of spin waves with finite wave vector observed here. The basic theory describing how the demagnetizing field modifies the FMR frequency of uniformly magnetized ellipsoidal bodies was developed by Kittel.[30] So, for an ellipsoid magnetized along its axis of rotational symmetry ($z$-axis), this spatially-uniform frequency is given by

$$f_{FMR} = \gamma\sqrt{(H + 4\pi(N_{xx} - N_{zz})M_S)(H + 4\pi(N_{yy} - N_{zz})M_S)}, \qquad (1)$$

where $H$ is the applied magnetic field, $\gamma$ is the gyromagnetic ratio, $M_S$ is the saturation magnetization, and $N_{\alpha\alpha}$ are the diagonal elements of the demagnetizing tensor. Here, we use Equation (1) to develop a method based on the ideas of Marchenko and Krivoruchko[31,32] by which to calculate the spatially dependent $f_{FMR}(r)$ in general non-ellipsoidal bodies undergoing uniform precession. Using the calculated ground micromagnetic state of the sample, we define a local curvilinear coordinate system as follows. The static magnetization direction defines the local $z$-axis. The other in-plane direction defines the local $y$-axis, and the $x$-axis is defined by $\hat{x} = \hat{z} \times \hat{y}$. Then, we apply modest uniform deflections to the magnetization along the local $x$- and $y$-axes (labelled $m_x'$ and $m_y'$ below) and calculate the resulting additional effective magnetic fields $h_x$ and $h_y$ numerically (using MuMax) and therefore the effective demagnetising tensor elements[33,34] $N_{xx}(r)$ and $N_{yy}(r)$ from

$$\begin{pmatrix} h_x \\ h_y \end{pmatrix} = -\begin{pmatrix} N_{xx} & N_{xy} \\ N_{yx} & N_{yy} \end{pmatrix}\begin{pmatrix} m_x' \\ 0 \end{pmatrix}, \quad \begin{pmatrix} h_x \\ h_y \end{pmatrix} = -\begin{pmatrix} N_{xx} & N_{xy} \\ N_{yx} & N_{yy} \end{pmatrix}\begin{pmatrix} 0 \\ m_y' \end{pmatrix}. \qquad (2)$$

The tensor element $N_{zz}(r)$ is obtained directly from the static effective field distribution. The off-diagonal components were found to be negligible here. So, only the diagonal components were used to determine $f_{FMR}(r)$ from Equation (1).

Physically, the local FMR frequency $f_{FMR}(r)$ defined by equations (1) and (2) represents the frequency at which the local susceptibility (i.e. one calculated neglecting the spatial dispersion) will peak at the given point (or region) of the sample. Alternatively, one could interpret it as the frequency of spin waves for which the point will serve as a classical "turning point", e.g. in a quasi-classical theory of wave propagation. In the turning point, the wavelength of the waves diverges, enabling coupling between the spin waves and the incident uniform microwave magnetic field at that frequency. A more detailed account of the method and the discussion of its advantages and limitations can be found in Ref. 35.

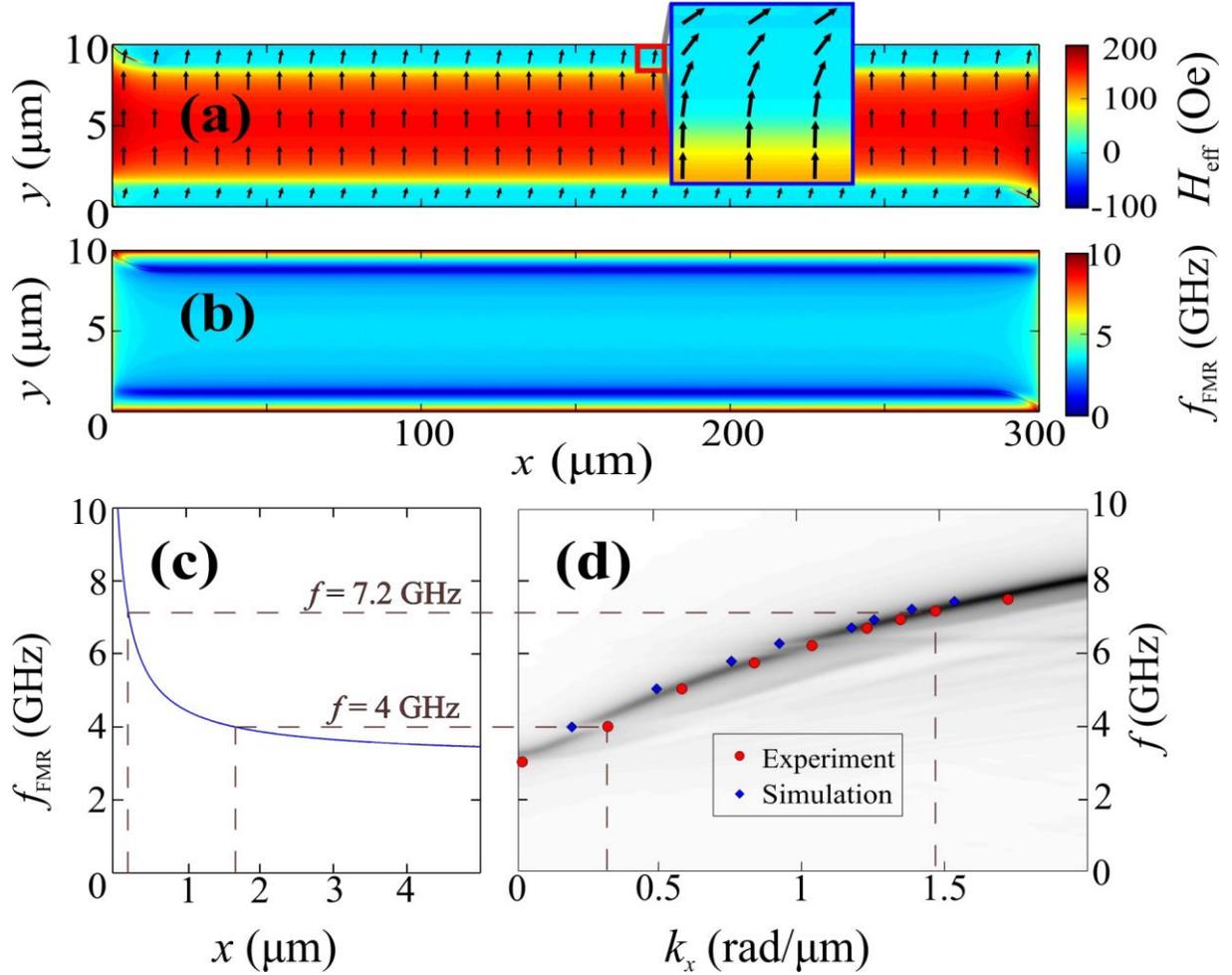

Fig. 4  (a) The calculated projection of the static effective field onto the magnetization in the ground magnetic state is shown with the color scale for the 100 nm thick Permalloy stripe of finite length. The unit vectors (averaged over 205 × 410 mesh cells) show the orientation of the relaxed magnetization. (b) The calculated spatial distribution of $f_{FMR}(r)$ calculated using Equation (1) is shown for the state from panel (a). (c) A cross-section of the $f_{FMR}(r)$ along the length of the stripe is shown for the first 5 μm from its left end and $y = 5$ μm. (d) The spin-wave dispersion calculated from the simulations of an infinitely long stripe is shown by grayscale, with the overlaid red circles and blue diamonds showing the points deduced from the measurements and simulations of the finite-length stripes, respectively. For the dispersion calculation, the results of the simulations were spatially smoothed so as to mimic the experimental resolution. The horizontal dashed lines extending from panel (c) to (d) illustrate the correspondence between the source region and wave number of propagating spin waves excited at frequencies of 4 and 7.2 GHz.

The ground state micromagnetic configuration of the transversely-magnetized stripe is shown in Fig. 4 (a). This was calculated for the same stripe model as discussed in Supplementary information. As expected, the configuration is significantly non-uniform in terms of both the magnetization orientation and the effective field strength. The corresponding calculated distribution of the local frequency of resonance is shown in Fig. 4 (b). Close to all four edges of the stripe, we note an increase in $f_{FMR}$, as compared to that calculated at the center of the stripe. Fig. 4 (c) shows the $f_{FMR}$ profile along the long symmetry axis of the stripe, close to its left end. As one moves from the end, there is a steep and then gradual reduction in $f_{FMR}$ until the value of 3.2 GHz is reached at the center of the stripe. Hence, upon exciting the entire system at a greater frequency $f > 3.2$ GHz, a specific region of magnetization close to the stripe's edges is excited at resonance. As indicated by the horizontal dashed lines in Fig. 4 (c), excitation at frequencies of 4 GHz and 7.2 GHz, for example, drives in resonance the magnetization regions around the $x$ coordinates of 1.65 μm and 0.19 μm, respectively. This local excitation injects propagating spin waves into the adjacent region of magnetization with reduced $f_{FMR}$ and then the rest of the stripe. In the region of significantly non-uniform $f_{FMR}$ (within about 5 μm from the end of the stripe), the wavelength of the excited spin wave will adiabatically increase to satisfy the dispersion relation,[36] until a mostly uniform landscape of $f_{FMR}$ is reached. The final value of the wavelength of the spin wave excited at a particular frequency can be found from the dispersion relation shown in Fig. 4 (d). For example, the wavelength is 18.9 μm and 3.5 μm at the frequencies of 4 GHz and 7.2 GHz, respectively, which agrees well with the experimental results. As mentioned earlier, the mechanism of the spin-wave injection is similar to that from Ref. 10 except here the frequency at which the spin waves are excited is not fixed but can rather take on any value, as dictated by the driving field frequency. We find that a suitably non-uniform graded $f_{FMR}(r)$ distribution in a thin film element allows excitation of spin waves of any wavelength that is supported by the film and explains the broad tuneability of excited spin waves we observe.

The distribution of $f_{FMR}(r)$ is also notably non-uniform and non-monotonic along the $y$-axis, reflecting the competition between the static and dynamic components of the demagnetizing field (Fig. 5). The static demagnetizing field due to the curving magnetization close to the edges causes $f_{FMR}$ to reduce, but at distances even closer, the dynamic demagnetizing field dominates, leading to a significant and rapid increase in $f_{FMR}$. The latter increase is responsible for the driving field exciting spin waves propagating from the two long edges of the stripe, as observed in form of "stripes" parallel to the horizontal edges (in their immediate vicinity) in Kerr images from Figs. 2 and 3 (see also the supplementary movies) and also from Ref. 13. The excitation of such modes has been predicted theoretically,[37] although only standing modes have been resolved experimentally.[21] Once again, this highlights the added value of the concept of local FMR frequency introduced here compared to interpretations based solely on the spatial distribution of the static internal field.

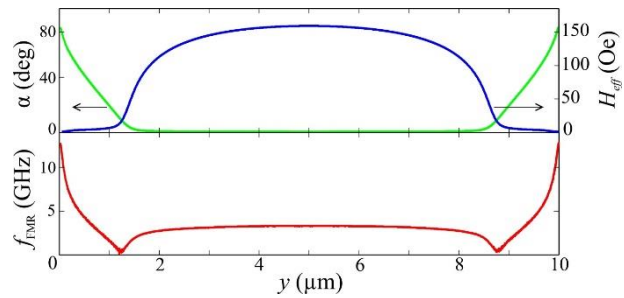

Fig. 5  The top panel shows the numerically simulated coordinate dependence of the angle $\alpha$ between the static magnetization and the *y*-axis and of the projection of the total internal field $H_{\text{eff}}$ on the direction of the static magnetization in the infinite stripe.  The bottom panel shows the corresponding distribution of the local FMR frequency $f_{\text{FMR}}$ calculated from the same simulations using equations (1) and (2).

In conclusion, we have demonstrated experimentally and numerically that magnetization at edges of a patterned ferromagnetic stripe serves as a source of propagating spin waves when driven by a uniform microwave magnetic field at frequencies above that of the bulk quasi-uniform mode.  The excitation arises from the spatial variation of the dynamic demagnetizing field intrinsic to the magnetic structure, rather than from spatial inhomogeneity of the excitation field.  The dynamic demagnetizing field increases locally the magnetic resonance frequency, thereby enabling the global microwave magnetic field to couple to specific regions of magnetization.  To explain our findings, we have developed a simple model by which to calculate the local resonance frequency across non-uniform magnetic configurations.  This excitation scheme is general – here, we have used the dynamic demagnetizing field to raise the local frequency of resonance, but in principle any magnetic non-uniformity could be used.  The diffraction-limited resolution of the TRSKM setup has restricted the investigation to spin waves with micrometer wavelengths, while it has been shown numerically that the scheme of spin-wave emission is highly scalable.[38]  Hence, our findings demonstrate an alternative approach devoid of the restrictions associated with nanoscale electrical circuitry by which magnonic technology[1,2] could be realized.

**Supplementary material**

See supplementary material for details of the experimental data fittings and of the numerical simulations.  The data associated with our results are available from http://hdl.handle.net/10871/28522.

**Acknowledgements**

The research leading to these results has received funding from the Engineering and Physical Sciences Research Council (EPSRC) of the United Kingdom (Project Nos. EP/L019876/1, EP/L020696 and EP/P505526/1).

# Supplementary information

**Supplementary note 1. Fitting of the Kerr signals used to obtain the spin-wave wavelength and decay length**

To deduce the wavelength $\lambda$ and decay length $l$ of the experimentally observed propagating spin waves, each Kerr signal $S(x)$ was fitted to an exponentially decaying harmonic wave function, given by

$$S(x) = S_0 \cdot \sin\left(\frac{2\pi}{\lambda}x + \varphi_0\right) \cdot \exp\left[-\frac{x}{l}\right] \cdot \left(1 + erf\left(\frac{x-a}{b}\right)\right) + cx + d, \quad (S1)$$

where $S_0$ and $\varphi_0$ are the spin-wave amplitude and phase, $a$ and $b$ are parameters of the standard error function, and $c$ and $d$ represent a linear background (Fig. S1.1). To decrease the noise level and thereby to improve the fit quality, the signals were averaged across the middle part of the stripe width ($y = 5 \pm 2.5$ µm). This fit procedure was implemented for each measured frequency, and the deduced $\lambda(f)$ and $l(f)$ dependences are shown in Fig. S1.2 together with the ratio $l(f) / \lambda(f)$.

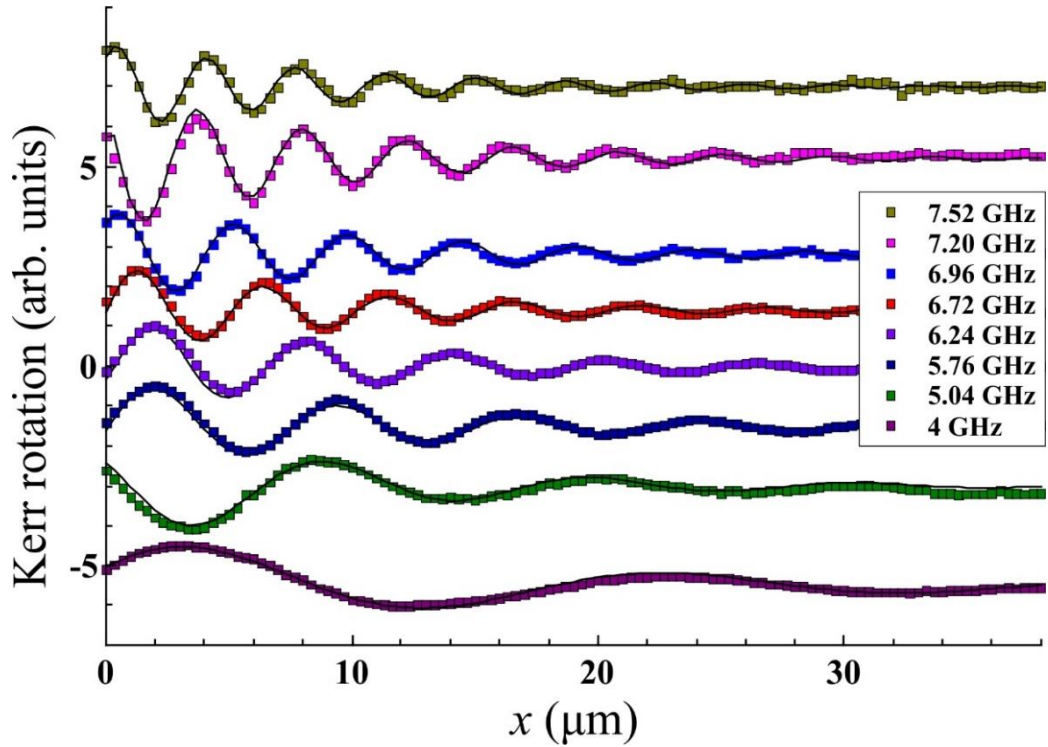

Fig. S1.1. The measured Kerr profiles along the long symmetry axis of the stripe ($x$ axis) are shown by symbols for different frequencies of the continuous-wave excitation. The solid lines are best fits to Equation (S1).

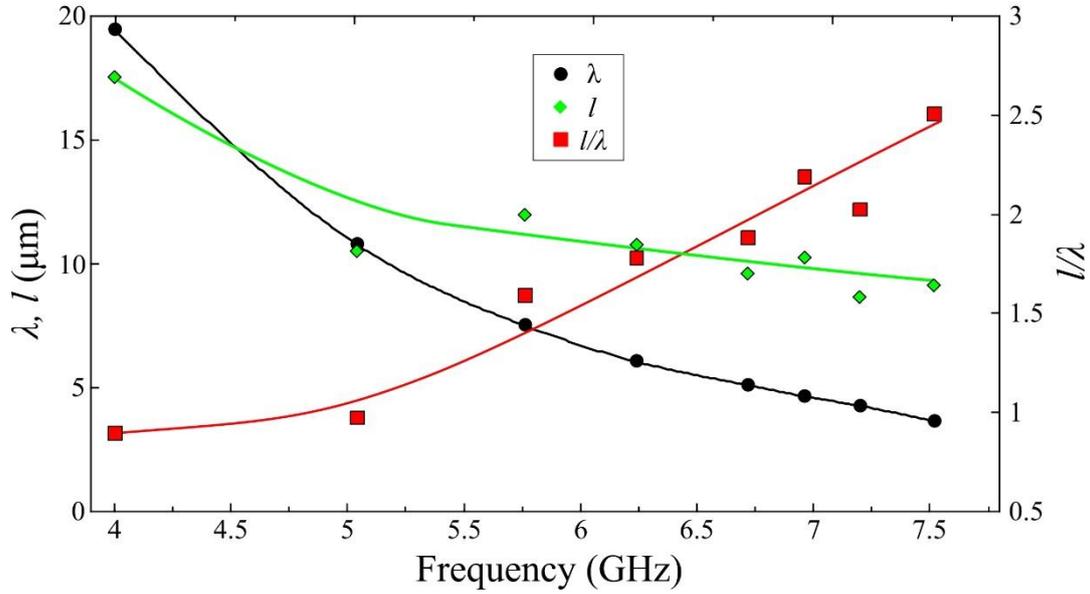

Fig. S1.2. The fitted values of the wavelength $\lambda$, decay length $l$ and their ratio $l/\lambda$ for the different values of the frequency of the continuous-wave excitation are shown by symbols. The solid lines are guides for the eye.

**Supplementary note 2. Micromagnetic simulations**

To calculate the dispersion of the spin waves that are supported by the stripe, micromagnetic simulations were performed using MuMax.[39] The stripe was modelled as a ferromagnetic slab 409.6×10×0.1 μm$^3$, meshed with 400×312.5×12.5 nm$^3$ cells, and assumed to have the Permalloy-like magnetic properties. The best agreement with the experiment was found for the saturation magnetization $M_s$ = 700 G, the exchange stiffness $A$ = 1.3 μerg/cm, the gyromagnetic ratio $\gamma/2\pi$ = 2.8 MHz / Oe, and the Gilbert damping parameter $\alpha$ = 0.008. Periodic boundary conditions were applied along the length of the stripe, so that the simulated stripe was effectively infinite. In the first stage of the simulations, only the uniform bias magnetic field $H$ = 200 Oe was applied along the short axis of the stripe, and a small additional magnetic field of 2 Oe was applied along the stripe's length to remove the degeneracy of the micromagnetic states. The ground magnetic state of the magnetization across the stripe was then calculated. In the second stage of the simulations, spin waves were excited using a short pulse of magnetic field of the form

$$h_z(x,t) = A \frac{\sin[k_c(x-x_0)]}{k_c(x-x_0)} \cdot \frac{\sin[2\pi f_c(t-t_0)]}{2\pi f_c(t-t_0)} ,  \quad (S2)$$

where $A = 10$ Oe was the pulse amplitude, and $k_c = 7$ rad µm$^{-1}$ and $f_c = 15$ GHz were the wave number and frequency "cut-offs", respectively; $x_0 = 204.8$ µm and $t_0 = 200$ ps were the pulse offsets in space and time, respectively. The spatial distributions of magnetization were recorded every 20 ps for a total duration of 40.96 ns. To obtain the spin-wave dispersion, a two-dimensional Fast-Fourier Transformation (FFT) was applied to the calculated $M_z$ (the out-of-plane projection of magnetization), as a function of time and the $x$-coordinate. The amplitudes of the complex Fourier coefficients were then summed across the middle part of stripe width ($y = 5\pm2.5$ µm) to yield the spectral density (SD)

$$SD(k_x, f) = \sum_{y,z} |FFT[M_z(x,y,z,t)]| \quad (S3)$$

To verify the proposed mechanism of spin-wave emission, a micromagnetic simulation of a finite length stripe was performed. The stripe was modelled as a ferromagnetic slab $300\times10\times0.1$ µm$^3$, mesh size was $4096\times2048\times1$ cells. The material parameters were kept the same as above. The stripe was relaxed to a ground state and excited with a uniform external magnetic field given by

$$h_z = A\sin(2\pi ft) \cdot \exp(-20/ft) , \quad (S4)$$

where A = 10 Oe and $f$ are the amplitude and frequency of exciting magnetic field, respectively. The second term in Eq. (S4) provides a "soft start" and prevents excitation of higher-frequency precessional modes. The simulations were repeated for different frequencies ranging from 4 to 7.52 GHz. Each simulation lasted as long as required for the higher-frequency modes to decay and for the steady-state harmonic oscillation of magnetization to be achieved. To determine the wavelength $\lambda$ of the emitted spin waves, the out-of-plane component of the magnetization $M_z$ for the last time frame was smoothed using a Gaussian filter with a width equal to the resolution limit of our experimental setup ~ 400 nm, and the FFT was applied to $M_z$ along the $x$-axis of the stripe. The wavelength was then extracted from the position of maximum in the calculated Fourier spectrum. The extracted values $f(k)$ are shown by symbols in Fig. 4 (d).

[1] V. V. Kruglyak, S. O. Demokritov, and D. Grundler, "Magnonics", *J. Phys. D: Appl. Phys.* **43**, 26030 (2010).

[2] S. A. Nikitov, D. V. Kalyabin, I. V. Lisenkov, A. Slavin, Y. N. Barabanenkov, S. A. Osokin, A. V. Sadovnikov, E. N. Beginin, M. A. Morozova, Y. P. Sharaevsky, *et al*, "Magnonics: a new research area in spintronics and spin wave electronics", *Phys. Uspekhi* **58**, 1002 (2015).

[3] A. K. Ganguly and D. C. Webb, "Microstrip excitation of magnetostatic surface waves: theory and experiment", *IEEE. Trans. Microw. Theory Tech.* **23**, 998 (1975).

[4] V. Vlaminck and M. Bailleul, "Spin-wave transduction at the submicrometer scale: Experiment and modeling", *Phys. Rev. B* **81**, 014425 (2010).

[5] P. Gruszecki, M. Kasprzak, A. E. Serebryannikov, M. Krawczyk and W. Śmigaj, "Microwave excitation of spin wave beams in thin ferromagnetic films", *Sci. Rep.* **6**, 22367 (2016).

[6] Y. Au, M. Dvornik, T. Davison, E. Ahmad, P. S. Keatley, A. Vansteenkiste, B. Van Waeyenberge, and V. V. Kruglyak, "Direct excitation of propagating spin waves by focused ultrashort optical pulses", *Phys. Rev. Lett.* **110**, 097201 (2013).

[7] K.-J. Lee, A. Deac, O. Redon, J.-P. Nozieres, and B. Dieny, "Excitations of incoherent spin-waves due to spin-transfer torque", *Nature Mat.* **3**, 877 (2004).

[8] S. O. Demokritov, V. E. Demidov, O. Dzyapko, G. A. Melkov, A. A. Serga, B. Hillebrands, and A. N. Slavin, "Bose-Einstein condensation of quasi-equilibrium magnons at room temperature under pumping", *Nature* **443**, 430 (2006).

[9] E. Schlömann, "Generation of spin waves in nonuniform magnetic field. I. Conversion of electromagnetic power into spin-wave power and vice versa", *J. Appl. Phys.* **35**, 159 (1964).

[10] Y. Au, T. Davison, E. Ahmad, P. S. Keatley, R. J. Hicken, and V. V. Kruglyak, "Excitation of propagating spin waves with global uniform microwave fields", *Appl. Phys. Lett.* **98**, 122506 (2011).

[11] Y. Au, E. Ahmad, O. Dmytriiev, M. Dvornik, T. Davison, and V. V. Kruglyak, "Resonant microwave-to-spin-wave transducer", *Appl. Phys. Lett.* **100**, 182404 (2012).

[12] M. Arikan, Y. Au, G. Vasile, S. Ingvarsson, and V. V. Kruglyak, "Broadband injection and scattering of spin waves in lossy width-modulated magnonic crystal waveguides", *J. Phys. D.: Appl. Phys.* **46**, 135003 (2013).

[13] C. S. Davies, A. Francis, A. V. Sadovnikov, S. V. Chertopalov, M. T. Bryan, S. V. Grishin, D. A. Allwood, Yu. P. Sharaevskii, S. A. Nikitov, and V. V. Kruglyak, "Towards graded-

index magnonics: Steering spin waves in magnonic networks", *Phys. Rev. B* **92**, 020408 (2015).

[14] S. Wintz, V. Tiberkevich, M. Weigand, J. Raabe, J. Lindner, A. Erbe, A. Slavin, and J. Fassbender, "Magnetic vortex cores as tunable spin-wave emitters", *Nature Nanotechn.* **11**, 948 (2016).

[15] U. Fano, "Effects of configuration interaction on intensities and phase shifts", *Phys. Rev.* **124**, 1866 (1961).

[16] A. E. Miroshnichenko, S. Flach, and Y. S. Kivshar, "Fano resonances in nanoscale structures", *Rev. Mod. Phys.* **82**, 2257 (2010).

[17] Y. Au, M. Dvornik, O. Dmytriiev, and V. V. Kruglyak, "Nanoscale spin wave valve and phase shifter", *Appl. Phys. Lett.* **100**, 172408 (2012).

[18] W. Strauss, "Magnetoelastic waves in yttrium iron garnet", *J. Appl. Phys.* **36**, 118 (1965).

[19] B. A. Auld, J. H. Collins, and D. C. Webb, "Excitation of magnetoelastic waves in YIG delay lines", *J. Appl. Phys.* **39**, 1598 (1968).

[20] P. E. Zilberman, A. G. Temiryazev, and M. P. Tikhomirova, "Excitation and dispersion of exchange spin waves in iron-yttrium garnet films", Zh. Eksp. Teor. Fiz. **108**, 281 (1995).

[21] J. P. Park, P. Eames, D. M. Engebretson, J. Berezovsky, and P. A. Crowell, "Spatially resolved dynamics of localized spin-wave modes in ferromagnetic wires", *Phys. Rev. Lett.* **89**, 277201 (2002).

[22] A. Barman, V. V. Kruglyak, R. J. Hicken, J. M. Rowe, A. Kundrotaite, J. Scott, and M. Rahman, "Imaging the dephasing of spin wave modes in a square thin film magnetic element", *Phys. Rev. B* **69**, 174426 (2004).

[23] C. Bayer, J. Jorzick, B. Hillebrands, S. O. Demokritov, R. Kouba, R. Bozinoski, A. N. Slavin, K. Y. Guslienko, D. V. Berkov, N. L. Gorn, and M. P. Kostylev, "Spin-wave excitations in finite rectangular elements of $Ni_{80}Fe_{20}$", *Phys. Rev. B* **72**, 064427 (2005).

[24] R. D. McMichael and B. B. Maranville, "Edge saturation fields and dynamic edge modes in ideal and nonideal magnetic film edges", *Phys. Rev. B* **74**, 024424 (2006).

[25] I. Lisenkov, V. Tyberkevych, S. Nikitov, and A. Slavin, "Theoretical formalism for collective spin-wave edge excitations in arrays of dipolarly interacting magnetic nanodots", *Phys. Rev. B* **93**, 214441 (2016).

[26] R. W. Damon and J. R. Eshbach, "Surface magnetostatic modes and surface spin waves", *Phys. Rev.* **118**, 5 (1960).

[27] S. O. Demokritov (ed.) "Spin wave confinement" (Pan Stanford Publishing, Singapore, 2009).

[28] K. Yu. Guslienko, S. O. Demokritov, B. Hillebrands and A. N. Slavin, "Effective dipolar boundary conditions for dynamic magnetization in thin magnetic stripes", *Phys. Rev. B* **66**, 132402 (2002).

[29] K. Yu. Guslienko and A. N. Slavin, "Boundary conditions for magnetization in magnetic nanoelements", *Phys. Rev. B* **72**, 014463 (2005).

[30] C. Kittel, "Ferromagnetic resonance", *J. Phys. et radum* **12**, 332 (1951).

[31] A. N. Marchenko and V. N. Krivoruchko, "Magnetic structure and resonance properties of a hexagonal lattice of antidots", *Low. Temp. Phys.* **38**, 157 (2012).

[32] V. N. Krivoruchko and A. I. Marchenko, "Spatial confinement of ferromagnetic resonances in honeycomb antidot lattices", *J. Magn. Magn. Mater.* **324**, 387 (2012).

[33] J. R. McDonald, "Ferromagnetic resonance and the internal field in ferromagnetic materials", *Proc. Phys. Soc. A* **64**, 968 (1951).

[34] N. Vukadinovic, "High-frequency response of nanostructured magnetic materials", *J. Magn. Magn. Mater.* **321**, 2074 (2009).

[35] C. S. Davies, V. D. Poimanov, and V. V. Kruglyak, "Mapping the magnonic landscape in patterned magnetic structures", (2009), arXiv:1706.03212.

[36] V. E. Demidov, M. P. Kostylev, K. Rott, J. Münchenberger, G. Reiss, and S. O. Demokritov, "Excitation of short-wavelength spin waves in magnonic waveguides", *Appl. Phys. Lett.* **99**, 082507 (2011).

[37] C. Bayer, J. P. Park, H. Wang, M. Yan, C. E. Campbell, and P. A. Crowell, "Spin waves in an inhomogeneously magnetized stripe", *Phys. Rev. B* **69**, 134401 (2004).

[38] C. S. Davies and V. V. Kruglyak, "Generation of propagating spin waves from edges of magnetic nanostructures pumped by uniform microwave magnetic field", *IEEE Trans. Magn.* **52**, 2300504 (2016).

[39] A. Vansteenkiste, J. Leliaert, M. Dvornik, M. Helsen, F. Garcia-Sanchez, and B. Van Waeyenberge, "The design and verification of MuMax3", *AIP Advances* **4**, 107133 (2014).